\documentclass[12pt,preprint]{aastex}

\shorttitle{Cold Dust in Kepler's Supernova Remnant}
\shortauthors{Morgan et al.}

\begin{document}

\title{Cold Dust in Kepler's Supernova Remnant.}

\author{H.L.Morgan\altaffilmark{1},
L.Dunne\altaffilmark{1}, S.A.Eales\altaffilmark{1},
R.J.Ivison\altaffilmark{2}, M.G.Edmunds\altaffilmark{1}}

\altaffiltext{1}{Department of Physics and Astronomy, University of Wales Cardiff PO Box 913, CF24 3YB; \\haley.morgan@astro.cf.ac.uk}
\altaffiltext{2}{Astronomy Technology Centre, Royal
Observatory, Blackford Hill, Edinburgh EH9 3HJ}

\begin{abstract}

The timescales to replenish dust from the cool, dense winds of
Asymptotic Giant Branch stars are believed to be greater than the
timescales for dust destruction.  In high redshift galaxies, this
problem is further compounded as the stars take longer than the age of
the Universe to evolve into the dust production stages.  To explain
these discrepancies, dust formation in supernovae (SNe) is required to
be an important process but until very recently dust in supernova
remnants has only been detected in very small quantities.  We
present the first submillimeter observations of cold dust in Kepler's
supernova remnant (SNR) using SCUBA.  A two component dust temperature model
is required to fit the Spectral Energy Distribution (SED) with
$T_{warm} \sim 102$K and $T_{cold} \sim 17$K. The total mass of dust
implied for Kepler is $\sim 1M_{\odot}$ - 1000 times greater than
previous estimates.  Thus SNe, or their progenitors may be important
dust formation sites.
\end{abstract}


\section{Introduction} \label{intro}

Recent blank field submillimeter (submm) surveys and observations of
distant quasars have discovered a population of extremely dusty
objects at high redshifts (Smail, Ivison \& Blain 1997; Bertoldi et
al. 2003) implying that large quantities of dust are present in the
Universe at $z > 4$ (Eales et al. 2003; Ivison et al. 2003, Issak et
al 2002).  Dust is generally believed to form in the cool atmospheres
of giant stars and expelled via slow dense winds, yet at these
redshifts, the time needed for such a process is greater than the age
of the Universe.  Morgan \& Edmunds (2003) constructed models of the
maximum dust mass produced in high redshift galaxies.  Using a simple
method to estimate the condensation efficiency of dust in Stellar
Winds (SW's) they found there is not enough time at redshifts $>$ 4
for significant amounts of dust to be made in SW's unless star
formation rates are {\it{very}} high and dust destruction rates are
low.  Dust production in SNe however provides an explanation for the
dust seen in high redshift SCUBA sources as SNe evolve on a much
shorter timescale.

Theoretical models of nucleation of dust grains in supernova
gas suggest condensation efficiencies of $\sim 0.1$ (Clayton, Deneault
\& Meyer 2001) and $\sim 0.2-0.8$ (Todini \& Ferrara 2001) could be
reached in Type II SNe regardless of initial metallicity. However, the
observational evidence that SNe produce large quantities of dust is
very weak.  Infra-Red (IR) observations of young SNRs with IRAS and
ISO found only $\sim 10^{-3}$ and $10^{-4}M_{\odot}$ of dust in Tycho
and Kepler (Douvion et al. 2001b) and $10^{-4}M_{\odot}$ in Cas A
(Douvion, Lagage \& Pantin 2001a; Arendt, Dwek \& Moseley 1999).
Observations of SN1987A have found a similar amount of dust (Dwek et
al. 1992; Wooden 1997) although this could be much higher if the
ejecta is clumpy (Lucy et al. 1991).  One possible reason for the
discrepancy between observations and theory would be if the dust in
the SNRs is cold and virtually impossible to detect at IR wavelengths
but observable in the submm waveband.  Recent SCUBA
observations of Cas A (Dunne et al. 2003) have shown that a colder
population of grains does exist with an estimated $2 - 4M_{\odot}$ of
dust, which implies a condensation efficiency of $0.5 - 0.8$ for a
30$M_{\odot}$ progenitor star.  In this Letter, we present the first
submm observations of another young Galactic SNR, Kepler's SNR.

Kepler's SNR is the remnant from the early 1600's (Green 2001).
Recent estimates place it at a distance of 4.8 - 6.4kpc (Reynoso \&
Goss 1999) so we adopt 5kpc.  The progenitor and type of Kepler's SN
are still hotly debated.  It was originally thought to be a Type Ia SN
due to its light curve, and its distance from the Galactic plane
suggests the progenitor was an old population II object.  However,
slow moving optical knots present in the region around the SNR show
substantial overabundance of nitrogen, which can be most naturally
explained by the the CNO cycle in massive stars (Bandiera 1987;
Borkowski, Blondin \& Sarazin 1992).  The slow expansion velocity and
high densities of the knots suggest they were not ejected in the SN
explosion but rather in an earlier stellar wind, which again suggests
a massive star progenitor (Blair, Long \& Vancura 1991).  X-Ray models
of the remnant provide further evidence for a dense medium. These
require ambient densities of a few H atoms per cubic centimeter
(Kinugasa \& Tsunemi 1991), which is 100 times greater than the
ambient density of the interstellar medium (ISM) expected at 600pc
above the Galactic plane (Whittet 2001). All these observations
indicate there was dense circumstellar material (CSM) predating the
SNR, which is most easily explained by the stellar winds from a
massive star.  For these reasons, Kepler's supernova has been
classified as a Type Ib SN.  Bandiera (1987) has proposed that the
progenitor of Kepler's SNR was a massive star which was ejected from
the galactic plane.  In this model, the star moves through its own
stellar wind, compressing the wind material into a bow shock in the
direction of motion, and this produces the N-S asymmetry seen in the
X-Ray and radio observations.

\section{Data Reduction} \label{kepdatared}

We observed Kepler's SNR with SCUBA at the 15m James Clerk Maxwell
Telescope (JCMT) on Mauna Kea in Hawaii.  SCUBA simultaneously
observes at 450 and 850$\mu$m using bolometer arrays of 91 and 37
bolometers respectively.  The field of view (fov) is $\sim 2.3$
arcmin, but is slightly smaller at 450$\mu$m.  The beam size (FWHM) at
these wavelengths is $\sim 8$ and 15 arcsecs respectively.  As
Kepler's SNR is actually larger than the fov of SCUBA, we observed the
remnant with six 'jiggle-map' observations, each jiggle map being
centered around the SNR with a chop throw of 180 arcsec. 

The data was reduced using the SURF software package (Sandell et al.
2001).  We made corrections for atmospheric absorption using the
opacities derived from skydip measurements with the JCMT and
from polynomial fits to the sky opacity measured by the Caltech
Submillimeter Observatory at 225GHz.  Noisy bolometers were flagged
and large spikes were removed.  Any noise correlated across the array
(i.e sky noise) was removed using the SURF program {\small{REMSKY}}.
A map was made from the six individual datasets using one-arcsec
pixels.  We created noise maps at each wavelength using the Monte-Carlo
technique described in Eales et al. (2000).  The S/N images at 850 and
450$\mu$m are shown in Figure~\ref{fig-1} with the 850$\mu$m map
clearly showing a ring like morphology similar to the X-Ray and radio
observations.  The 450$\mu$m map shows emission in both the
north-western and south-eastern quadrant.  The noise at 450$\mu$m has
a much stronger dependence on the atmospheric conditions than the
noise at 850$\mu$m.  The conditions were much worse when we observed
the other two quadrants, so the lack of obvious emission in those
quadrants and also in the centre of the remnant may partly be the
result of the increased noise in these regions.  

The image was calibrated using primary calibrators which were observed
with similar chop throw to the observations of the SNR.  The flux was
measured in eight apertures, which were placed on the maps to include
all the obvious emission from the SNR (using the S/N maps as a guide).
The flux calibration was carried out using apertures of the same size
on the calibrator map.  The errors in the fluxes were estimated as
described in Dunne et al. (2000) with calibration errors of 15$\%$ and
20$\%$ respectively.  The fluxes measured over the whole remnant at
850 and 450$\mu$m are $1.0\pm 0.2$Jy and
$3.0 \pm 0.7$Jy respectively.  Thus the ratio of 450 to 850$\mu$m flux
for the SNR as a whole is 3.

\section{Estimating the Dust Mass}
\label{sec:dust}

\begin{figure}
\rotatebox{270}{
\epsscale{0.6}
\plotone{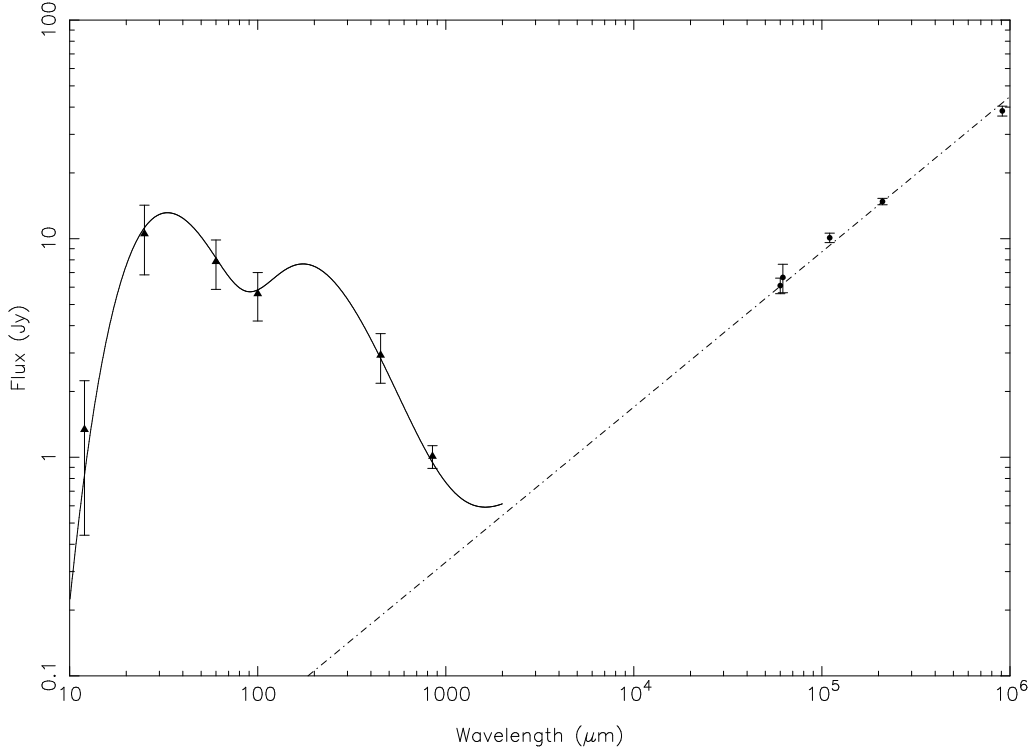}}
\caption{Kepler's IR - Radio SED for the  
IRAS 12, 25, 60 and 100$\mu$m fluxes, the 850 and 450$\mu$m
measurements from this work and the radio (Matsui et al. 1984; Delaney
et al. 2002).  The solid line represents a two temperature fit to the
data points with $\beta = 1.2$, $T_w = 102K$, $T_c = 17K$. The straight
line through the radio measurements is a power law with a spectral
index -0.71.  The 100$\mu$m flux was chosen here to be
5Jy rather than the lower value of
$\sim$3Jy observed by Arendt (1989) and Braun (1987).  This was a
conservative choice as using the lower value would produced a more
pronounced second peak in the SED.  To obtain the best fit greybody
through these points would then require a cold temperature of $\sim
11K$ and hence more than double the derived dust mass.  \label{fig-2}}
\end{figure}

Supernova remnants are strong sources of synchrotron radiation which can
contaminate the longer submm wavelengths.  To obtain the submm flux
which is emitted by dust grains only, we need to correct for this
contamination.  Using the average of the IR fluxes from the literature
(Arendt 1989; Saken, Fesen \& Shull 1992; Braun 1997) and the submm
fluxes from this work, we can produce Kepler's IR-Radio SED
(Figure~\ref{fig-2}).  This clearly shows that the emission from dust
is in excess of the synchrotron emission.  We can estimate the
contribution of the synchrotron to the submm using the five radio
measurements which are well fitted by a power law slope $S_{\nu}
\propto \nu^{\alpha}$ where $\alpha=-0.71$.  
Hence the synchrotron emission is responsible for
30$\%$ of the submm flux at 850$\mu$m and $6\%$ at 450$\mu$m.   

\begin{figure}
\plotone{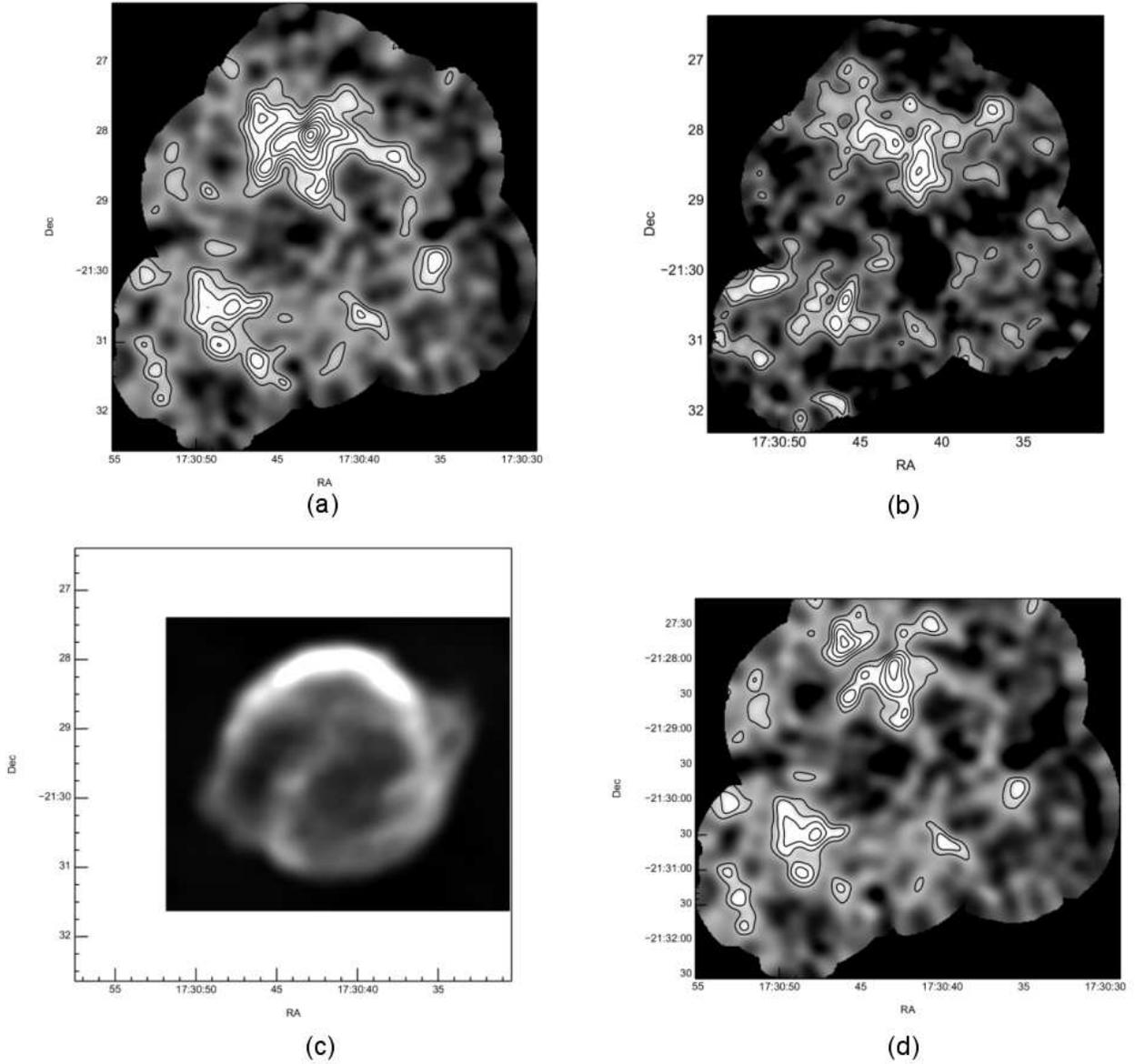}
\caption{Signal/Noise maps of the Kepler SNR. (a)
850$\mu$m smoothed to $14''$ with contours 3$\sigma$, 4$\sigma$,
5$\sigma$...etc.  (b) 450$\mu$m image smoothed to
the 850$\mu$m resolution with contours $1.5\sigma$, $2.5\sigma$,
$3.5\sigma$...  (c) 6cm radio image of Kepler (Delaney et al. 2002)
smoothed to the resolution of the 850$\mu$m beam.  (d)
The synchrotron subtracted 850$\mu$m image with levels showing 3$\sigma$,
4$\sigma$, 5$\sigma$... \label{fig-1}}
\end{figure}

Comparison of the 450 and 850$\mu$m maps with a radio map also clearly
indicates that there is submillimetre emission in excess of the
expected synchrotron emission.  Figure~\ref{fig-1}(c) shows the 5GHz
VLA image of the SNR (Delaney, private communication).  The map shows
the well known N-S asymmetry associated with the interaction with a
dense CSM.  The 850$\mu$m map also shows a N-S asymmetry but is much less
pronounced than the radio.  The 450$\mu$m map shows an even lesser
N-S asymmetry.  A point-by-point comparison of the maps also
indicates excess submm emission.  The strong emission visible in the
south west on the submm maps is at a position where there is hardly
any emission on the radio map.

To correct the submm images for the contribution
of synchrotron radiation we smoothed the VLA radio image to the
resolution of the submm images and scaled it to the submm wavelengths
using the power law spectrum shown in Figure~\ref{fig-2}.  We then
subtracted it from the submm image.  The synchrotron-subtracted
850$\mu$m image is shown in Figure~\ref{fig-1}(d).  The brighter
northern limb in the 850$\mu$m unsubtracted image has been removed and
the 450 and 850$\mu$m images now follow a similar morphology with two
bright emission regions in the north and south of the remnant.  The
peaks on the 450 and 850$\mu$m maps do not exactly coincide with each
other but this is not unexpected given the low S/N at 450$\mu$m and
the fact that we are extrapolating a long way in frequency from 5GHz
to the submm waveband.  The remnant is known to have variations in the
spectral index of order $\pm 15\%$ (Delaney et al. 2002).  This may
produce errors in the synchrotron-corrected maps although the general
agreement between the 450 and 850$\mu$m images suggests this is not a
problem.  Note also that the fact that the ratio of 450 to 850$\mu$m
is 3 makes it practically impossible to explain the excess
submm emission by some synchrotron process.  

The dust mass can be measured directly from the flux at submm
wavelengths using $M_d={S_{\nu}D^2 \over{\kappa_{\nu}B_{\nu}(\nu ,
T)}}$ (Hildebrand, 1983) where $S_{\nu}$ is the flux density measured
at frequency $\nu$, $D$ is the distance and $\kappa_{\nu}$ is the dust
mass absorption coefficient.  $B$ is the Planck function and T is the
dust temperature.  We tried fitting a single temperature greybody to
the SED ($S_{\nu} \propto B_{\nu}\nu^{\beta}$) but we could not obtain
an acceptable fit.  So we fitted a two-temperature greybody allowing
$\beta$ and the warm and cold temperatures to vary.  The best fit SED
gave $\beta=1.2$, $T_{w} = 102K$ and $T_{c} = 17K$.  The possible
reasons why the dust could be so cold are discussed in Dunne et al (2003)
and will be considered in future work (Dunne et al, in prep).  We
cannot rule out the presence of dust existing at intermediate
temperatures between these two extremes until further observations
around the peak of the SED ($200\mu$m) are made.  We used a bootstrap
technique to derive errors on these values.  We created 3000 sets of
artificial fluxes from the original data and then fit our two
temperature model to each set.  We derived our errors from the
distribution of $T_{w}$, $T_{c}$ and $\beta$ produced by these fits.

The largest uncertainty in the dust mass comes from the uncertainty in
$\kappa$.  We have followed Dunne et al. (2003) in trying three
different values of $\kappa$ from the literature: (1) $\kappa_{850} =
0.6 - 1.1m^2kg^{-1}$, the range observed in laboratory studies of
clumpy aggregates.  (2) $\kappa_{850} = 0.16 - 0.8m^2kg^{-1}$ the
range observed in circumstellar environments/reflection nebulae and
(3) $\kappa_{850} = 0.04 - 0.15m^2kg^{-1}$, the range observed for the
diffuse ISM where dust is likely to have encountered extensive
processing.  In each case we used the average values of the ranges to
estimate the dust mass which are given in Table~\ref{tabl-1}.  In Cas
A, the higher $\kappa$ values were required to give a reasonable dust
mass.  Using the highest $\kappa$ values for Kepler gives a lower
range of $0.2 - 0.3 M_{\odot}$ whereas using the lowest $\kappa$
values gives $1 - 3M_{\odot}$.  Whichever value of $\kappa$ we use we
obtain at least three orders of magnitude more dust than obtained from
mid-IR measurements.  This is an order of magnitude lower than the
dust observed in Cas A but the far greater brightness of Cas A than
Kepler at all wavelengths implies a (as yet not understood) difference
in the dynamics of the explosions.

\section{Discussion}

Kepler's SNR still appears to be in an early evolutionary
phase in which the swept up mass is not much more than the total
ejected mass (Dickel 1987; Hughes 1999).  If the interstellar density
at the location of Kepler's SNR is $n_H \sim 0.001 - 0.1cm^{-3}$
(Bandiera 1987; Borkowski et al. 1992) then given the size of the
remnant and assuming a normal gas to dust ratio of 160, the maximum
dust mass swept up by the SNR is $\sim 10^{-3}M_{\odot}$.


Evidence for an even denser CSM comes from slow moving optical knots
observed outside of the ejecta.  These are thought to originate from
accelerated clumps of circumstellar material and have densities
$n_H\sim 1000cm^{-3}$, which imply a pre-shock density of $100cm^{-3}$
(Blair et al. 1991).  If this density is representative of the CSM
surrounding the entire remnant, then the blast wave could have swept
up $100M_{\odot}$ of gas and a dust mass of $\sim 1M_{\odot}$.  This
is however, at odds with the mass observed in the X-Ray which suggests
that there is only $\sim 2 - 5M_{\odot}$ of gas within the SNR
(Kinugasa \& Tsunemi 1991; Borkowski et al. 1992).

Some additional arguments that the dust has not been swept up come
from considering the morphologies of the different images.  Some of
the dust is found between the forward and reverse shocks of the X-Ray
gas, where the gas is thought to be dominated by the ejecta rather
than the swept up material (Hughes 1999).  A large fraction of the
dust is also found in the south of the SNR, where the radio image
shows no evidence for a strong interaction between the SNR and a dense
CSM.  However, some of the submm emission is found outside the X-Ray
and Radio remnants.  This may be due to some of the dust being formed
in the progenitor's stellar wind or due to the velocity of the ejecta
moving faster in these regions.  The lack of a detailed position for
the two shock fronts does not allow for a complete comparison though
further work will try to address this problem (Dunne et al, in prep).

Could the dust be from a massive star, formed in its pre-supernova RSG
and WR phases?  The maximum observed dust mass lost in WR winds is $6
\times 10^{-7}M_{\odot}yr^{-1}$ (Marchenko et al. 2002).  The violent
mass loss from the WR will blow the material outwards into a shell
with outer radius depending on the velocity and duration of the wind.
If the wind velocity is $\sim 1000kms^{-1}$ and if the WR wind was
responsible for the dust seen at the radius of the submm emission, the
WR phase of the precursor must have only lasted around $10^{4}$
years. This would produce only $0.01M_{\odot}$ of dust.  The
interaction between the previous RSG and WR winds may have created
more dust but the maximum yield of heavy elements ejected from the
most massive stars during the RSG phase is $< 2 M_{\odot}$.  The
condensation efficiency in these atmospheres was estimated to be $<
10\%$ (Morgan \& Edmunds 2003) and hence it is difficult to produce
all of the observed dust in this way.  Note, if the condensation
efficiency is indeed higher, this {\it could} explain the lower end of
the cold dust mass in Kepler and hence SNe progenitors would be
responsible for polluting the high redshift galaxies.


The mass of dust we observe is lower but similar to the mass of dust
found in Cas A.  If Kepler's progenitor was a 30$M_{\odot}$ star
(Bandiera 1987), the condensation efficiency could have been $\sim 0.1
- 0.6$.  If we assume that the dust in Kepler's SNR is representative
of the dust formed in every 30$M_{\odot}$ progenitor then this is
roughly comparable with the dust mass/SNe given in Todini \& Ferrara
(2001).  Using their models with the condensation efficiency obtained
for Kepler's SNR from this work, we estimate that SNe inject $\sim 5 -
15(\times 10^{-3})M_{\odot}yr^{-1}$ into the ISM.  If the dust
injection rate for giant stars is $\sim 5(\times
10^{-3})M_{\odot}yr^{-1}$ (Whittet 2001; Morgan \& Edmunds 2003) then
SNe are injecting around 1 - 3 times more dust than stellar winds.
Adding supernova or massive stars as a source of dust to the Galactic
interstellar medium explains the high redshift submm observations
and could well resolve the current discreprancy (Jones et al. 1994)
between dust production and destruction rates.

\acknowledgements

HM would like to acknowledge a Cardiff University studentship and LD
is supported by a PPARC postdoctoral fellowship.  We would also like
to thank Tracey Delaney for providing the radio image and Eli Dwek for useful
discussions.  We thank the referee for invaluable comments on the draft
mauscript.

\clearpage

\begin{table}
\begin{center}
\caption{The parameters from the SED fitting
using minimum $\chi^2$ fit to the points.  The dust masses are
the median of the distribution from the bootstrap technique (same as
the best fit values) with
68$\%$ confidence interval quoted as the error. The values of $\kappa$
are (as in Dunne et al. 2003) $\kappa_1(450)=1.5m^2kg^{-1}$,
$\kappa_1(850)=0.76m^2kg^{-1}$ (laboratory studies);
$\kappa_2(450)=0.88m^2kg^{-1}$, $\kappa_2(850)=0.3m^2kg^{-1}$
(observations of evolved stars); $\kappa_3(450)=0.26m^2kg^{-1}$,
$\kappa_3(850)=0.07m^2kg^{-1}$ (observations of the diffuse
ISM). \label{tabl-1}}
\begin{tabular}{cccccc} \tableline
\tableline
\multicolumn{3}{c}{Parameters} &\multicolumn{3}{c}{Dust Mass
($M_{\odot}$)}\\ \tableline
\multicolumn{1}{c}{$\beta$}&\multicolumn{1}{c}{$T_h$
(K)}&\multicolumn{1}{c}{$T_c (K)$} &\multicolumn{1}{c}{$\kappa_1$}&
\multicolumn{1}{c}{$\kappa_2$} & \multicolumn{1}{c}{$\kappa_3$}\\
\tableline
\\
$1.2 \pm 0.4$ & $102\pm 12$ & 17$^{+3}_{-2}$ & $M_{450}$ =
$0.3^{+0.2}_{-0.1}$ & 0.5$^{+0.3}_{-02}$ & 1.7$^{+0.9}_{-0.61}$\\
& & & & &\\
& &  & $M_{850}$ = $0.3\pm 0.1$ & 0.6$\pm 0.2$ &
2.7$^{+0.6}_{-0.5}$\\
\\
\tableline
\tableline
\end{tabular}
\end{center}
\end{table}

\clearpage

\clearpage

\end{document}